\documentclass[%
 reprint,
 amsmath,amssymb,
 aps,
]{revtex4-1}

\usepackage{graphicx}
\usepackage{dcolumn}
\usepackage{bm}
\usepackage{color}
\usepackage[colorlinks, citecolor=blue,urlcolor=black,linkcolor=black]{hyperref}
\usepackage{amsthm}
\usepackage{enumitem}

\begin{document}

\preprint{APS/123-QED}

\title{Self-Organizing Quantum Networks}

\author{Dong Jiang}
\author{Wei-cong Huang}
\author{Chao-hui Gao}
\author{Jia Liu}
\email{Email address: jialiu@nju.edu.cn}
\author{Li-jun Chen}
 \email{Email address: chenlj@nju.edu.cn}
\affiliation{%
State Key Laboratory for Novel Software Technology, Nanjing University, Nanjing, 210046, P.R.China
}%

\begin{abstract}
As the inevitable development trend of quantum key distribution, quantum networks have attracted extensive attention, and many prototypes have been deployed over recent years.
Existing quantum networks based on optical fibers or quantum satellites can realize the metro and even global communications.
However, for some application scenarios that require emergency or temporary communications, such networks cannot meet the requirements of rapid deployment, low cost, and high mobility.
To solve this problem, we introduce an important concept of classical networks, i.e., self-organization, to quantum networks, and give two simple network prototypes based on acquisition, tracking, and pointing systems.
In these networks, the users only need to deploy the network nodes, which will rapidly, automatically, and adaptively organize and mange a quantum network.
Our method expands the application scope of quantum networks, and gives a new approach for the design and implementation of quantum networks.
It also provides users with a low-cost access network solution, and can be used to fundamentally solve the  security problem of classical self-organizing networks.
\end{abstract}


\maketitle

\section{Introduction}
Quantum Key Distribution (QKD), invented by Bennett and Brassard in 1984 \cite{bennett1984quantum}, is a technique for distributing a shared key to two distant users, Alice and Bob, such that the key is perfectly secure from the eavesdropper, Eve.
Since QKD offers proven unconditional security guaranteed by the fundamental laws of quantum mechanics, it has attracted intensive study and many advancements have been achieved over recent years \cite{scarani2009security}.
However, QKD is just a point-to-point secure communication solution, obviously, its development trend is quantum network, which has received increasing attention, and many prototypes have been implemented around the world \cite{ritter2012elementary}.
Generally, existing quantum networks can be divided into two categories: one is the optical fiber based quantum networks, e.g., DARPA deployed in United States \cite{elliott2005current}, SECOQC  demonstrated in Austria \cite{peev2009secoqc}, and the inter-city quantum networks implemented in China \cite{chen2010metropolitan}.
Recently, with the successful launch of the Micius quantum satellite, another type of quantum network, i.e., satellite based quantum network, has been realized \cite{liao2018satellite}.

Quantum networks based on optical fibers or quantum satellites can be used to realize metro and even global quantum communications.
Nevertheless, such networks need to lay optical fibers, launch quantum satellites, and build base stations.
For some application scenarios, such as geological exploration, battle field command, etc., this inevitably leads to the following problems:
first,  the users are often located in remote areas with complex terrain.
It is difficult to lay optical fibers or build base stations.
Second, the users need to quickly establish the network at low cost.
The time and cost of laying optical fibers or building base stations are unacceptable.
Third, the users may frequently change their locations.
It is unrealistic to modify or redeploy the network infrastructure once a user changes his location.
In practical applications, there are many similar scenarios.
They urgently need a solution that can be quickly deployed, at low cost, and support mobile users. However, there are few related studies and no systems that can be deployed immediately.

To meet above requirements, we introduce the concept of self-organization to quantum networks. Self-organization is an automatic technique originating from classical networks. It can rapidly, automatically, and adaptively realize the organization, configuration, management, optimization, and healing of mobile networks. Since Self-Organizing Networks (SON),  e.g., wireless sensor networks \cite{rawat2014wireless}, mobile \emph{ad-hoc} networks \cite{conti2014mobile}, and wireless mesh networks \cite{huang2015software}, lack the complexities of infrastructure setup and administration, the users are free to move, and can rapidly create or join a network anytime, anywhere. Clearly, such technology is completely suitable for solving above problems. Therefore, we follow the design pattern of SON, and propose two types of Self-Organizing Quantum Networks (SOQN) based on Acquisition, Tracking, and Pointing (ATP) systems \cite{yin2012quantum}.

The first network is Peer-to-Peer (P2P) SOQN, in which all network nodes are equally privileged. The second network is Client/Server (C/S) SOQN, which consists of clients and servers. In these networks, the users only need to deploy the network nodes, which will automatically and adaptively establish the optical links, organize the network, distribute the secret keys, and transmit the secret messages. The proposed solutions significantly increase the application areas of quantum networks, and give a new way for designing and implementing quantum networks. The users can also use our solutions to join existing optical fiber or quantum satellite based quantum networks at low cost, or to fundamentally solve the security problem of classical SON. Next we give a brief review of ATP system, followed by presenting the details of our networks.

\section{Acquisition, Tracking, and Pointing System}

\begin{figure}[!thpb]
	\centering
	\includegraphics[width=0.5\textwidth]{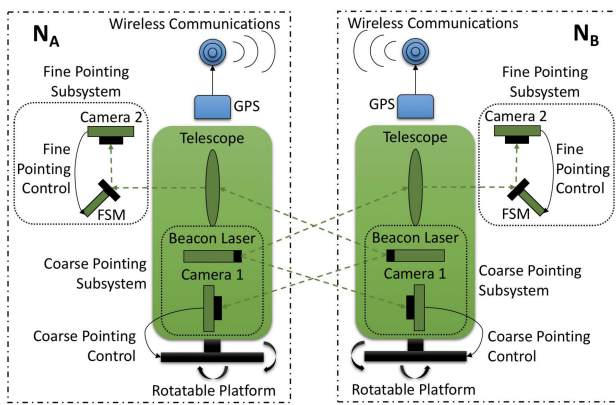}
	\caption{Establish an optical link between $N_A$ and $N_B$. FSM: Fast Steering Mirror.}
	\label{Fig::ATP}
\end{figure}

ATP system is used to automatically establish an accurate and stable optical link between two distant network nodes.
Since ATP is the key foundation of free-space laser communications, it has attracted intensive study, and many advanced solutions have been proposed over recent years \cite{bai2014predictive}.
In SOQN, ATP system of the Micius quantum satellite \cite{liao2017satellite} can be employed. It consists of two subsystems. One is called the coarse pointing subsystem, and the other is referred to as  the fine pointing subsystem. As shown in Fig. \ref{Fig::ATP},
when two users, say $N_A$ and $N_B$, want to establish an optical link, the detailed process is as follows:

\begin{itemize}[itemindent=0em]
	\item With the assistance of GPS and wireless communication circuits, nodes $N_A$ and $N_B$ acquire the latitudes, longitudes, and altitudes of their location, and broadcast their location information to each other.
	\item According to the received location information, $N_A$ and $N_B$ drive their rotatable platforms, and switch on the beacon lasers pointing to each other. When the beacon lights are spotted in the coarse cameras (camera 1), they drive their rotatable platforms to position the target beacon light at the center of the camera.
	\item After the beacon lights are spotted in the fine cameras (camera 2), they drive the fast steering mirrors to further correct the optical path. Different from the coarse pointing subsystem, the fine tracking subsystem shares the same optical path with the quantum communication system. Thus a high pointing accuracy can be achieved.
\end{itemize}

\section{Peer-to-Peer Self-Organizing Quantum Network}
\begin{figure}[!thpb]
	\centering
	\includegraphics[width=0.5\textwidth]{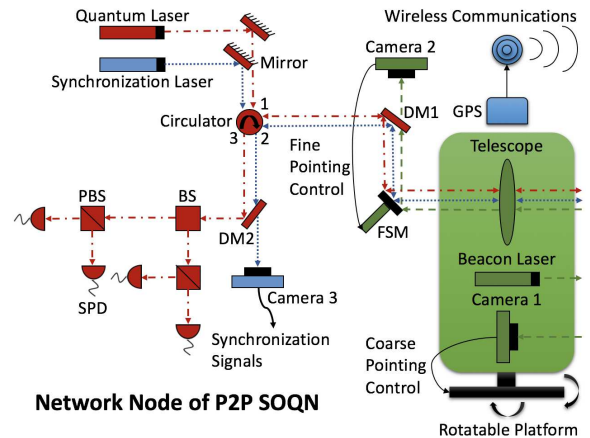}
	\caption{Network node of P2P SOQN. The circulator allows optical transmission from
		port 1 to 2 and from port 2 to 3. BS: Beam Splitter; DM: Dichroic Mirror; FSM: Fast Steering Mirror; PBS: Polarization Beam Splitter; SPD: Single Photon Detector.}
	\label{Fig::node}
\end{figure}

Now let us give a detailed description for P2P SOQN, which consists of equally privileged network peers or nodes.
As shown in Fig. \ref{Fig::node}, the node of P2P SOQN is composed of three parts: ATP system, quantum communication system, and classical communication system, which are marked with green, red, and blue, respectively.
These systems are used to establish optical links, distribute secret keys, transmit classical information and synchronization signals between two nodes, respectively.
In quantum communication system, the optical setup of BB84 protocol \cite{bennett1984quantum} is employed. The user can also use other QKD protocols, such as decoy state QKD \cite{lo2005decoy}, to distribute the secret keys.

\begin{figure}[!thpb]
	\centering
	\includegraphics[width=0.5\textwidth]{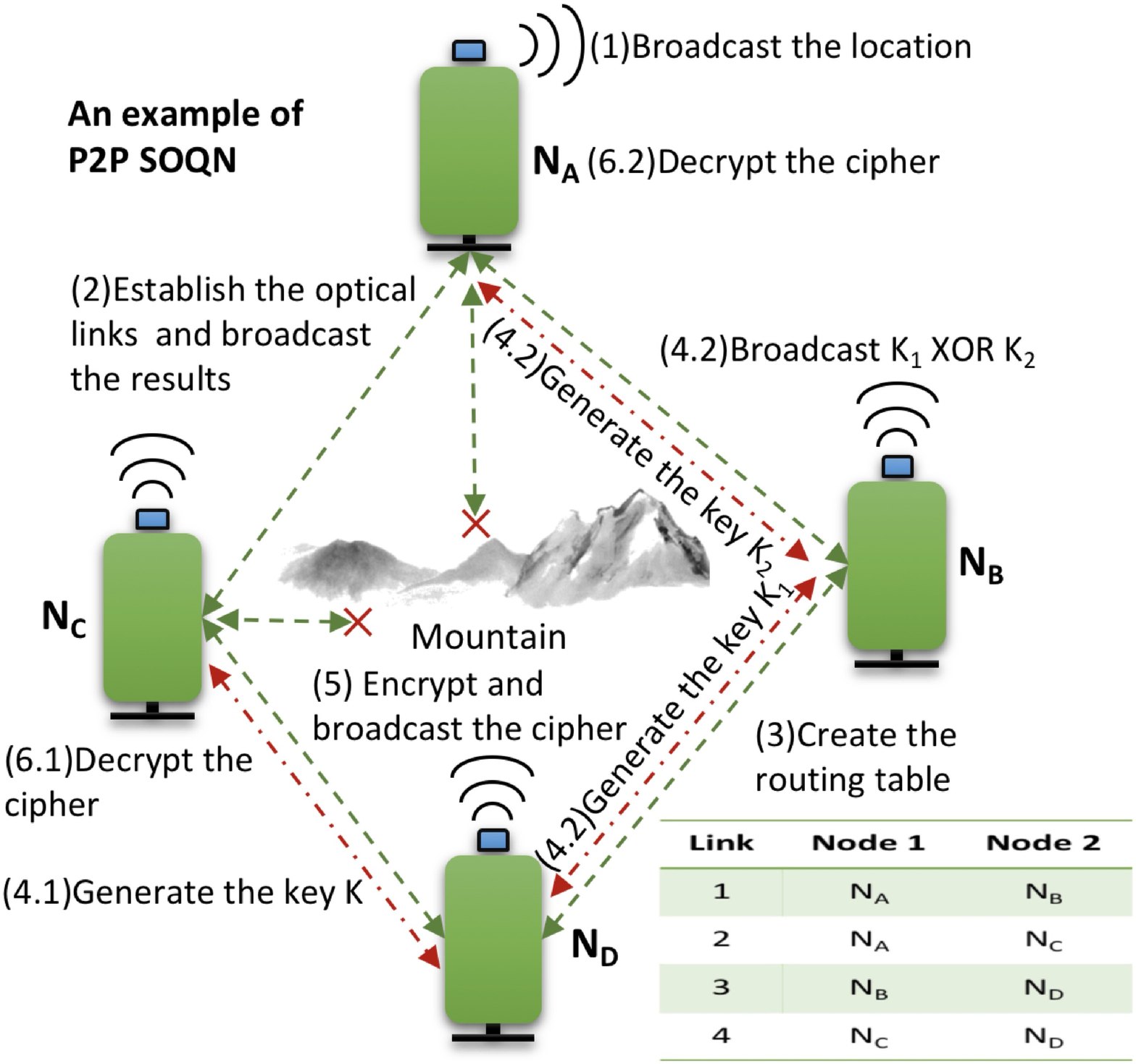}
	\caption{An example of P2P SOQN.}
	\label{Fig::P2PNetwork}
\end{figure}

As plotted in Fig. \ref{Fig::P2PNetwork}, after the users deploy their nodes. They can establish optical links and organize network as follows:

\begin{itemize}
	\item All nodes obtain and announce their location information.
	\item All nodes drive their ATP systems to establish optical links with each other, and broadcast the results.
	\item According to the results, all nodes create a routing table, which records all the valid optical links between the nodes.
	\item When two nodes want to generate a set of keys, they first query their routing tables.
	\begin{itemize}
		\item If there exists a link between them, they drive their ATP systems to point at each other, execute QKD protocol, and generate the key $K$.
		\item Otherwise, they find the relay node, which successively generates two set of keys $K_1$ and $K_2$ with the users, then broadcasts the results of $K_3=K_1\mathbin{\oplus}K_2$.
	\end{itemize}
	\item For the above two cases, the nodes use different methods to encrypt the plaintext $M$ and decrypt the ciphertext $C$.
	\begin{itemize}
		\item The sender encrypts the plaintext by $C=M\mathbin{\oplus}K$. The receiver decrypts the ciphertext by $M=C\mathbin{\oplus}K$.
		\item The sender encrypts the plaintext by $C=M\mathbin{\oplus}K_1$. The receiver decrypts the ciphertext by $M=C\mathbin{\oplus}K_2\mathbin{\oplus}K_3$.
	\end{itemize}
\end{itemize}

\begin{figure*}[!bthp]
	\centering
	\includegraphics[width=0.76\textwidth]{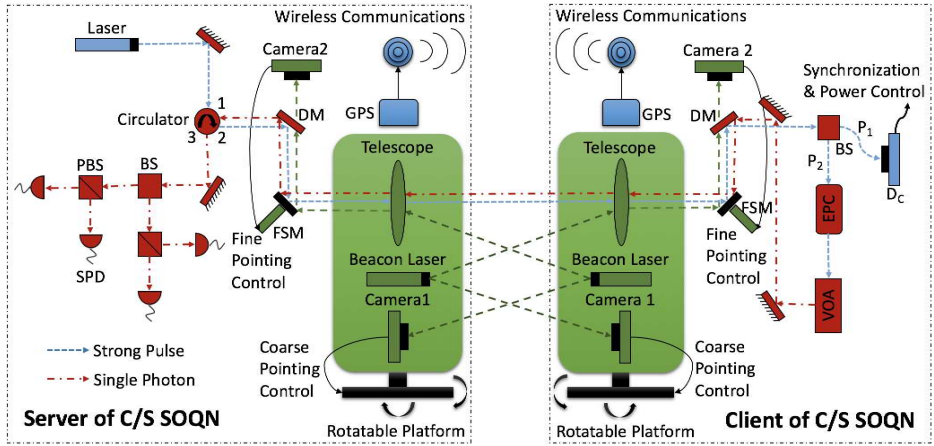}
	\caption{Server and client nodes of C/S SOQN. The circulator allows optical transmission from
		port 1 to 2 and from port 2 to 3. BS: Beam Splitter; DM: Dichroic Mirror; EPC: Electrical Polarization Controller; FSM: Faster Steering Mirror; PBS: Polarization Beam Splitter; SPD: Single Photon Detector; VOA: Variable Optical Attenuator.}
	\label{Fig::csnode}
\end{figure*}

\noindent In P2P SOQN, when a new user wants to join the network or a user needs to change his location, the node broadcasts a request after it is deployed. Then it tries to establish optical links with other nodes, and publishes the results. After all network nodes update their routing tables according to the published results, the node joins the network. Clearly, no matter the users want to establish or join a P2P SOQN, they only need to select a location and deploy their nodes, which can rapidly, automatically, and adaptively organize and manage the network.

\section{Client/Server Self-Organizing Quantum Network}

Next, we give a detailed description of C/S SOQN, which follows the design pattern of mobile or cellular networks \cite{asadi2014survey}. It consists of servers and clients that use two-way quantum communication protocol to transmit messages.
Generally, existing two-way quantum communication protocols can be divided into two categories. In the first category, e.g., quantum secure direct communication \cite{deng2004secure}, deterministic secure quantum communication \cite{jiang2017deterministic}, the receiver prepares and sends single photons. The sender encodes his messages on the received photons and sends them back.
In the second category, e.g., plug \& play QKD \cite{muller1997plug}, the receiver sends strong light pulses to the sender who encodes his messages on these pulses, attenuates them to single photon level, and sends them back. By measuring the received single photons, the receiver can obtain the sender's messages. Clearly, the second type of protocols are more suitable for C/S SOQN, because the single photons are only transmitted once over the free-space link (from the sender to the receiver), resulting in higher feasibility.
They are more similar to one-way QKD, which has been realized over 144km free-space link \cite{schmitt2007experimental}. Therefore, in C/S SOQN, the plug \& play QKD protocol is employed.

Many experimental demonstrations of plug \& play QKD use phase modulator to encode messages, which is obviously not applicable to free-space QKD. Therefore, we modify plug \& play protocol, using the polarization state as the message carrier.
The server and client nodes are plotted in Fig. \ref{Fig::csnode}. After the optical link is established, they can generate  keys as follows:

\begin{itemize}
	\item The server sends a strong light pulse to the client.
	\item When the pulse reaches the client node, it is separated into two pulses $P_1$ and $P_2$ by a 50/50 beam splitter. The detector $D_c$ measures the intensity of $P_1$, and triggers polarization controller to randomly prepare $P_2$ into one of the four states $|\phi\rangle\in\{|H\rangle, |V\rangle, |+\rangle, |-\rangle\}$, corresponding to horizontal, vertical, $+45^\circ$, $-45^\circ$ states, respectively. $P_2$ is then attenuated to the desired level by the variable attenuator according to the measurement result of $D_c$, and is sent back to the server.
	\item The server randomly selects a basis, measures the received state $|\phi\rangle$, and stores the measurement result as raw key.
	\item By conducting base sifting, error estimation, key reconciliation, and privacy amplification on the raw key, the server and the client can obtain a set final key.
\end{itemize}

\begin{figure}[!thpb]
	\centering
	\includegraphics[width=0.5\textwidth]{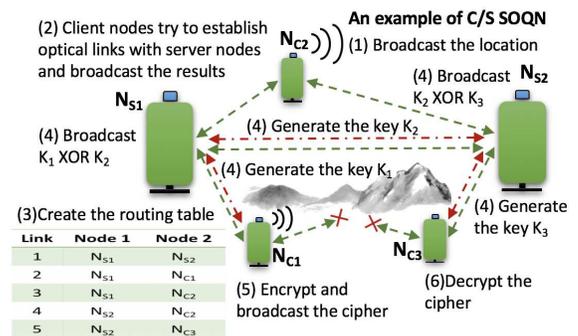}
	\caption{An example of C/S SOQN.}
	\label{Fig::csnetwork}
\end{figure}

Since the client node only provides encoding function, even though two users or clients are close to each other, they have to use one or more servers as relays to transmit messages. Thus, before the clients join the network, the servers must be deployed. These servers can automatically and adaptively organize a quantum network and create the routing table. Such network that only contains the servers is referred to as the backbone network. After the backbone network is established, as shown in Fig. \ref{Fig::csnetwork}, the clients can join the network as follows:

\begin{itemize}
	\item All clients and servers automatically obtain and broadcast their locations.
	\item  All clients drive their ATP systems, try to establish optical links with each server, and broadcast the results.
	\item According to the received results, all servers update their routing tables to record the valid optical links between the servers and the clients.
	\item When two clients want to transmit a message, they broadcast a request. The servers query their routing tables, selects serve nodes as relays to generate keys and transmit messages between clients (The process is similar to P2P SOQN, thus is not discussed in detail due to the limited space).
\end{itemize}
\noindent Similar to P2P SOQN, when a new client wants to join the backbone network, or a client has to change his location, he only needs to deploy his client node, which will automatically broadcasts a request. Then all servers announce their location. The client tries to establish optical links with all servers and broadcasts the results. After all servers update their routing tables according to the received results, the client joins the network.

\section{Discussion}

After introducing the details of the proposed networks, next, let us discuss their performances, including security, feasibility, and adaptiveness. First, the security of the protocols used in P2P and C/S SOQN have been proved in Ref. \cite{tomamichel2012tight} and Ref. \cite{beaudry2013seurity}, respectively. Although relay nodes may be used to transmit messages in SOQN, Ref. \cite{liao2018satellite} has realized satellite-relayed intercontinental quantum network, and shown that relay based quantum communications are secure. Besides, for plug \& play QKD, the eavesdropper can use Trojan-horse attacks to obtain the user's messages without leaving any trace. However, Refs. \cite{li2006improving,gisin2006trojan} prove that two-way QKD can resist such attacks by simply adding some additional optical elements and measuring part of received signals.
Second, compared with optical fiber based quantum networks, SOQN is environmentally sensitive, it may be affected by sunlight. However, experimental demonstrations have realized free-space QKD over 50 km in daylight \cite{liao2017long}. Therefore, our method is feasible with current technologies.
Third, as discussed above, in both P2P and C/S SOQN, no matter the users want to establish a network, or a new user wants to join the network, or a user has to change his location, all they need to do is deploy the network nodes, which can rapidly and automatically realize the network organization, key generation, and message transmission. SOQN achieves high adaptability, and thus completely meets the requirements of application scenarios that require emergency or temporary communications.

In addition, we know that the cost of a product significantly affects its application and promotion. For exiting quantum networks, if each user needs to lay an optical fiber or deploy a base station to join the network,  these networks cannot be widely used. In this paper, SOQN provides users with new access network solutions as follows: the quantum network operators deploy access points (P2P or server nodes) in their networks. By deploying the user nodes (P2P or client nodes), the users can join the network without laying optical fibers or deploying base stations. This method is similar to mobile networks. The networks are established and managed by the operators. The users only need to afford the cost of the network nodes, they can join and use the network at a very low cost. Thus such approach can significantly increase the possibility of commercialization of quantum networks.
Furthermore, with the development of computer science, especially quantum computing technologies, classical SON is facing increasing threats \cite{perrig2004security}. By using quantum method, SON can fundamentally solve the security problems.

\section{Conclusion}
To sum up, the concept of self-organization is introduced to quantum networks. And two types of SOQN, i.e., P2P and C/S SOQN, are proposed. In these networks, the users only need to deploy the network nodes, which will rapidly, automatically, and adaptively establish the optical links, organize the quantum network, distribute the secret keys, and transmit the secret messages. SOQN significantly expands the application scope of quantum networks. We hope our method can open up a new window for the study of quantum networks and classical SON.\\

\noindent\textbf{Funding.} The National Key Research and Development Program of China (No. 2017YFA0303700), the Major Program of National Natural Science Foundation of China (No. 11690030, 11690032), the National Natural Science Foundation of China (No. 61771236).\\

\noindent\textbf{Acknowledgement.} We thank Liang-liang Lu (School of Physics, Nanjing University, Nanjing,  P. R. China) for helpful discussion.

\end{document}